\documentclass{article}
\usepackage[main, final]{neurips_2025}

\usepackage[utf8]{inputenc}
\usepackage[T1]{fontenc}
\usepackage{hyperref}
\usepackage{url}
\usepackage{booktabs}
\usepackage{amsmath,amssymb}
\usepackage{nicefrac}
\usepackage{microtype}
\usepackage{xcolor}
\usepackage{graphicx}

\title{Ray-trax: Fast, Time-Dependent, and Differentiable Ray Tracing for\\
On-the-fly Radiative Transfer in Turbulent Astrophysical Flows}

\workshoptitle{Machine Learning and the Physical Sciences}

\author{
  \setlength{\tabcolsep}{2pt}
  
  \vspace{-1mm}
  Lorenzo Branca$^{1}$ \quad
  Rune Rost$^{1}$ \quad
  Tobias Buck$^{1}$\\[2pt]
  $^{1}$\footnotesize{Interdisciplinary Center for Scientific Computing, Heidelberg University, Germany}\\
  \footnotesize\texttt{rune.rost@stud.uni-heidelberg.de}\\
  \footnotesize\texttt{lorenzo.branca@iwr.uni-heidelberg.de}\\
  \footnotesize\texttt{tobias.buck@iwr.uni-heidelberg.de}
}

\begin{document}

\maketitle

\begin{abstract}
Radiative transfer is a key bottleneck in computational astrophysics: it is nonlocal, stiff, and tightly coupled to hydrodynamics.
We introduce \textbf{Ray-trax}, a GPU-oriented, fully differentiable 3D ray tracer written in JAX that solves the \emph{time-dependent} emission--absorption problem and runs directly on turbulent gas fields produced by hydrodynamic simulations.
The method favors the widely used on-the-fly \emph{emission--absorption approximation}, which is state of the art in many production hydro codes when scattering is isotropic.
Ray-trax vectorizes across rays and sources, supports arbitrarily many frequency bins without architectural changes, and exposes end-to-end gradients, making it straightforward to couple with differentiable hydro solvers while \emph{preserving differentiability}.
We validate against analytical solutions, demonstrate propagation in turbulent media, and perform a simple inverse problem via gradient-based optimization.
In practice, the memory footprint scales as $\mathcal{O}(N_{\text{src}}\,N_{\text{cells}})$ while remaining highly efficient on accelerators.
\end{abstract}

\section{Introduction}
Radiative feedback shapes star formation, galaxy evolution, and the circumgalactic/intergalactic medium \citep{1999MNRAS.305..449N, 2024MNRAS.527.8078O, 2025MNRAS.540.1745S}.
However, radiative transfer (RT) is often the step that limits resolution, cadence, or physics fidelity in large-scale simulations: it introduces global couplings, adds a fast characteristic speed ($c$), and requires angular and spectral resolution.
For many workflows, especially \emph{on-the-fly} coupling to hydrodynamics, an emission--absorption model (isotropic scattering) is the de facto standard \citep{2024FrASS..1146812W}; it captures photoheating and attenuation with a clean computational footprint and predictable scaling \citep{RHD1984}.
This paper presents \textbf{Ray-trax}, a compact JAX \citep{jax2018github} ray-tracing implementation designed around four goals:
(i) \emph{time dependence}, solving the initial-value radiative-transport problem with finite light-travel horizons;
(ii) \emph{GPU throughput}, by fully vectorizing the marching loop over rays and sources;
(iii) \emph{frequency-bin agnosticism}, treating the frequency dimension as a batch axis so users can add as many radiation bins as needed without changing code paths; and
(iv) \emph{differentiability}, enabling gradient-based inference and, crucially, the option to \emph{couple with differentiable hydro} while maintaining end-to-end gradients 
\url{https://github.com/lorenzobranca/Ray-trax.git}.

\section{Method and code structure}
\label{sec:method}

We model time-dependent, monochromatic transport of specific intensity
$I(\mathbf{x},\hat{\mathbf{n}},t)$ via
\begin{equation}
\frac{1}{c}\frac{\partial I}{\partial t} + \hat{\mathbf{n}}\!\cdot\!\nabla I
\;=\; -\,\kappa(\mathbf{x})\,I \;+\; j(\mathbf{x}) ,
\label{eq:rte_time}
\end{equation}
with space-only coefficients $\kappa$ and $j$ defined on a regular 3D grid.
Along characteristics we march in \emph{space} with step $\Delta s$ (so one
time step of size $\Delta t$ corresponds to a propagation horizon
$c\,\Delta t$). The code uses a semi-analytic/explicit update per substep
$k\!\to\!k{+}1$:
\begin{equation}
\mathbf{x}_{k+1}=\mathbf{x}_{k}+\hat{\mathbf{n}}\Delta s,\qquad
\tau_{k+1}=\tau_k+\kappa(\mathbf{x}_k)\,\Delta s,\qquad
I_{k+1}=I_k\,e^{-\kappa(\mathbf{x}_k)\Delta s}+j(\mathbf{x}_k)\,e^{-\tau_k}\,\Delta s,
\label{eq:update}
\end{equation}
i.e., exact attenuation over the step and a first-order quadrature for
emission. The number of substeps is
$N_s=\lceil c\,\Delta t/\Delta s\rceil$ (or a user-specified cap).
Despite the apparent simplicity of the approximation (isotropic scattering), it is state of the art in numerical simulations \citep{arepo-RT, ramses-RT, davide2020}. \\
We trace $N_\Omega$ nearly uniform directions on $\mathbb{S}^2$ using a
Fibonacci (golden-angle) lattice well suited to vectorization \citep{Gonzalez2010}. The per-ray
contributions are summed and scaled by $4\pi/N_\Omega$ to approximate the
solid-angle integral.\\
At each substep we read $j$ via trilinear interpolation \citep{HanssonSoderlund2022SDF} and (currently)
sample $\kappa$ with nearest-neighbor interpolation at the cell index used for the
previous position, then deposit the \emph{post-attenuation} intensity
$I_{k+1}$ back to the grid using the same trilinear weights. This makes the
discrete operator smooth in the voxel values and avoids scatter/gather
mismatch. Indexing is \emph{clamped} to the domain, so rays exiting the box
deposit to the nearest boundary voxels.\\
The single-source kernel
advances the field over one $\Delta t$ given
\texttt{(j\_map, kappa\_map, source\_pos, num\_rays, step\_size, radiation\_velocity, time\_step)}.
Multiple sources are supported by a simple loop-and-sum wrapper.

\paragraph{Vectorization, memory, and speed.}
Rays are independent conditioned on the medium, so we use
\texttt{vmap} over directions; the marching loop itself is written with unrolled
\texttt{lax.fori\_loop} and JIT compiled. On multi-GPU nodes we optionally
shard directions across devices with \texttt{NamedSharding} and perform a
final on-device sum per shard before reducing across shards; the code
requires $N_\Omega$ to be divisible by the device count.\\
The work scales as $\mathcal{O}(N_{\text{src}}\,N_\Omega\,N_s)$. In the
current implementation, each ray accumulates into a grid and these are
reduced at the end, so the transient memory is
$\mathcal{O}(N_\Omega\,N_{\text{cells}})$ (amortized by device sharding);
the returned field is one grid per time step (or the sum over sources in the
multi-source routine).\\
On a $128^3$ grid, tracing $4{,}096$ rays from $2{,}097$ sources across the full domain completes in $4.49\,\mathrm{s}$ (wall clock) on a single NVIDIA H200 GPU.

\paragraph{Frequency-bin agnosticism.}
Multi-frequency RT often forces architectural choices.
In Ray-trax the number of bins is a batch hyperparameter: adding bins simply increases parallel work.
No control-flow branches, no new kernels, and no changes to the differentiation rules are required.

\section{Benchmarks in uniform media and turbulent fields}
\label{sec:bench}

\begin{figure}[t]
\centering
\includegraphics[scale=0.45]{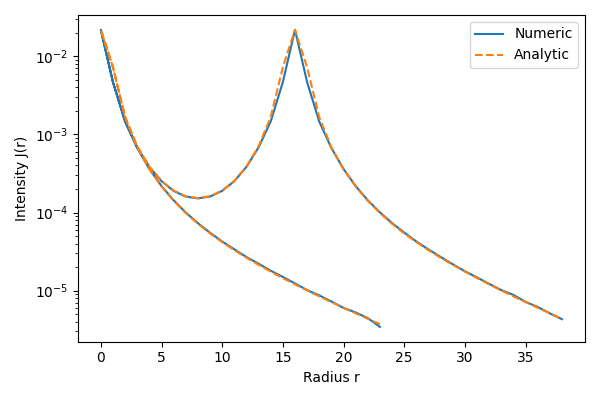}
\centering
\includegraphics[width=0.9\linewidth]{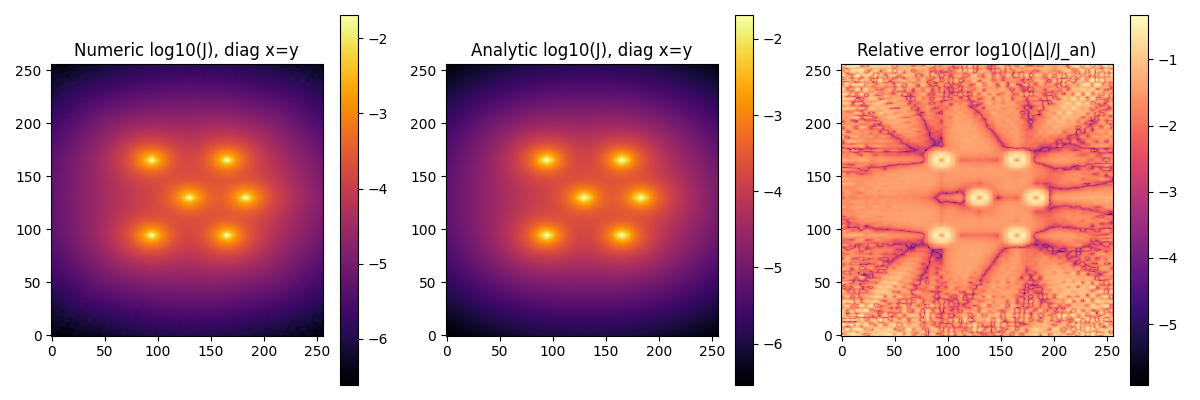}
\caption{Analytical benchmark in a uniform medium with multiple point sources. \textbf{Top:} Specific intensity $I$ along a line of sight through the domain center, shown for both $\pm$ directions; the sightline intersects a second source in one of the directions. \textbf{Bottom:} Diagonal slice of $I$ across the volume (left to right: numerical solution, analytical reference, and $\log_{10}$ residuals).}
\label{fig:analytic}
\end{figure}

\paragraph{Analytical validation.}
For constant $\kappa$ and point-like sources (represented numerically as narrow Gaussians) the reference intensity is
\begin{equation}
J_{\mathrm{ref}}(\mathbf{x}) \;=\; \sum_{i=1}^{N_{\text{src}}} \frac{L_i}{4\pi r_i^2}\,e^{-\kappa r_i},
\qquad r_i=\|\mathbf{x}-\mathbf{x}_i\|.
\label{eq:analytic}
\end{equation}
We compare $\log_{10} J$ slices (numeric vs.\ analytical) and relative-error maps, excluding the central voxel and a one-voxel border from metrics.
Convergence is monotonic in both the spatial step $\Delta s$ and the number of directions $N_\Omega$, as expected for first-order ray marching with uniform angular sampling. The average relative error with respect to the analytical solution is $3.16\%$.
A radial cut through a source validates the $1/r^2$ scaling and the exponential attenuation, shown in the top panel of Fig.~\ref{fig:analytic}.

\paragraph{Turbulent snapshots from hydro.}
We apply \texttt{Ray-trax} \emph{directly} to turbulent gas fields from hydrodynamic simulations. The opacity is prescribed as $\kappa(\mathbf{x}) \propto \rho(\mathbf{x})$, where $\rho$ is the turbulent density computed with the hydro code \texttt{jf1uid} \citep{2024arXiv241023093S}. Radiation sources are placed at the top $0.1\%$ density peaks. This setup is intended to mimic a typical turbulent, star-forming cloud.

\begin{figure}[t]
\centering
\includegraphics[width=0.91\linewidth]{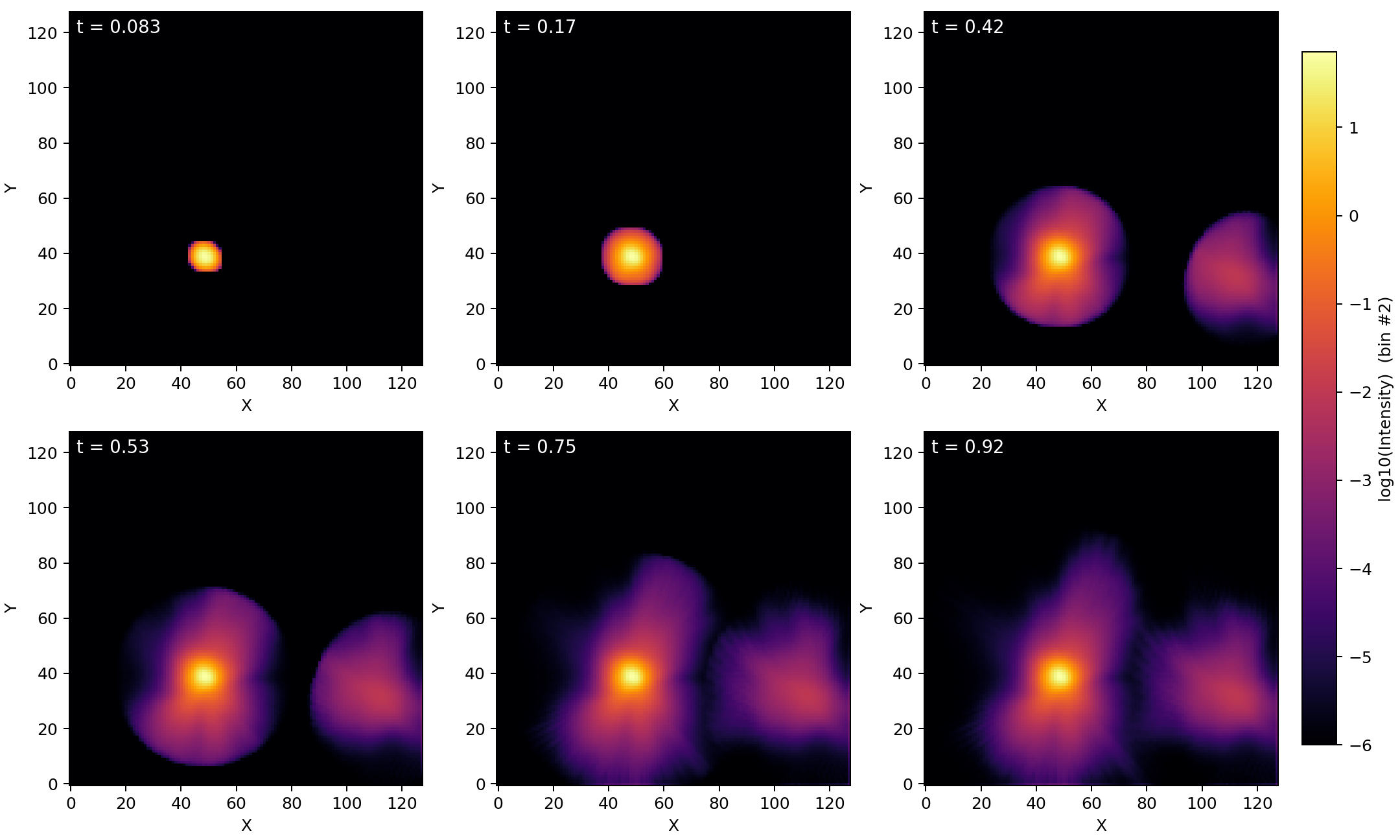}
\caption{Time-resolved propagation (from top left to bottom right) of specific intensity $I$ through a turbulent opacity field $\kappa(\mathbf{x})$ extracted from a hydrodynamic snapshot. The frequency bin is chosen to highlight shadowing. Time $t$ is in unit of total simulation time.}
\label{fig:turbulent}
\end{figure}

Figure~\ref{fig:turbulent} shows six $x$--$y$ slices of $\log_{10} I$ evolving over time: the morphology displays shadowing, porous channels, and illuminated bubbles consistent with the underlying flow structures.
For time-resolved visualizations we increase the light-travel horizon step by step to reveal the front progression; for strict time integration we fix $\Delta t$ and iterate.


\paragraph{On-the-fly suitability.}
In many production hydro codes, the \textbf{emission--absorption} approximation is the \emph{state of the art} for on-the-fly RT: it captures attenuation and local heating without introducing angular coupling from scattering.
\texttt{Ray-trax} is tailored to this use case: the operator is parallel over rays, frequency bins are batched, and the memory footprint is predictable and linear in $N_{\text{src}}$ and $N_{\text{cells}}$.
This makes it practical to insert RT substeps between hydro updates without derailing wall-clock budgets.

\section{Differentiability: inversion and coupling}
\label{sec:diff}

\paragraph{Gradient-based inversion.}
To illustrate end-to-end gradients we recover a scalar source amplitude $A$ with a fixed spatial template $e_0(\mathbf{x})$ from a reference field $I_{\mathrm{ref}}$.
We minimize either a plain MSE or a relative MSE using gradients $\partial \mathcal{L}/\partial A$ from \texttt{jacfwd}.
The 1D loss landscape $A\mapsto\mathcal{L}(A)$ is convex; the gradient solution matches the corresponding closed-form minimizer (under the chosen weighting).
This toy example readily extends to a small vector of parameters (e.g., a global $\kappa$ scale and a few amplitudes), and to multi-bin fits where each bin contributes an additive term in the loss.

\begin{figure}[t]
\centering
\includegraphics[scale=0.45]{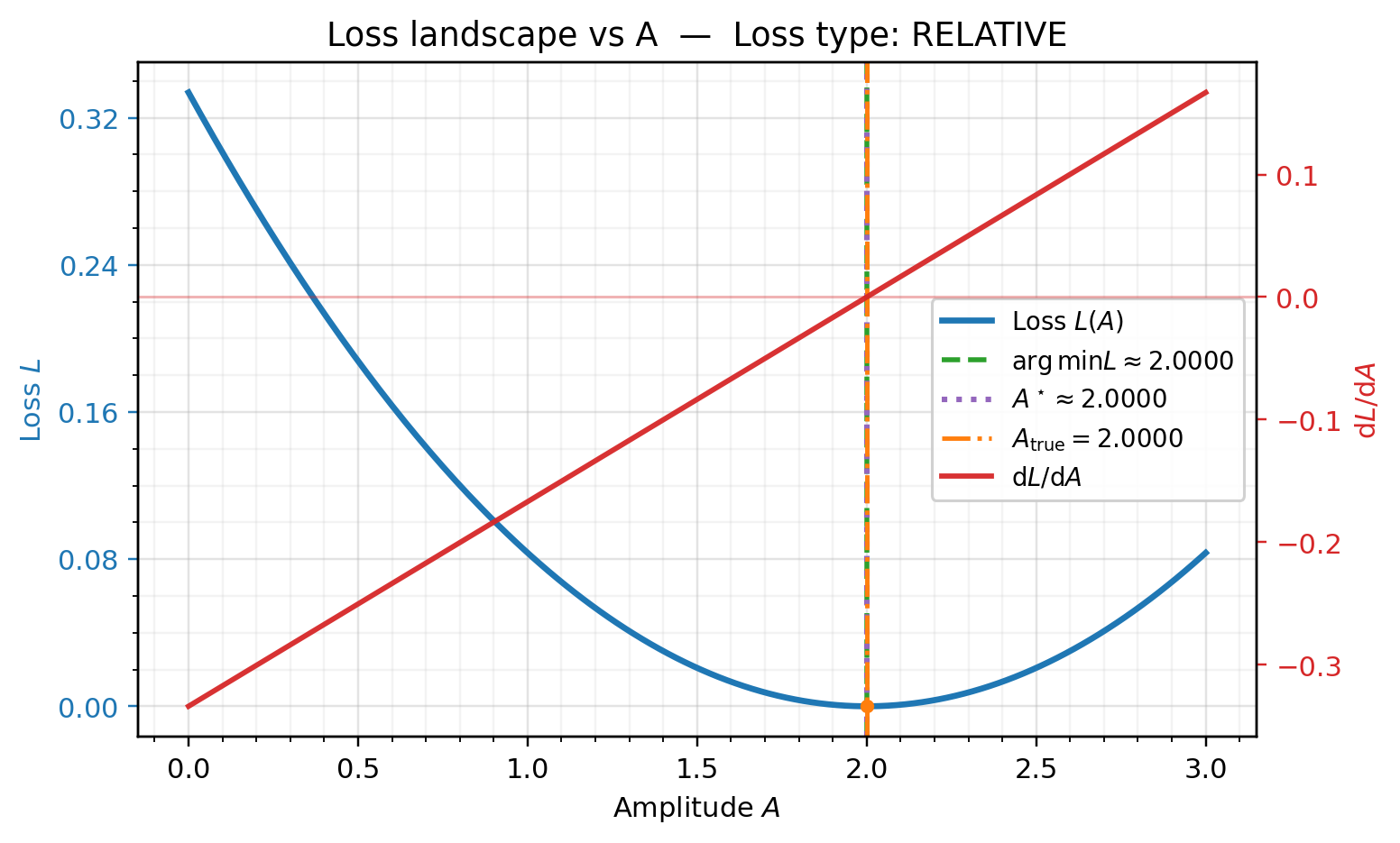}
\caption{End-to-end differentiability enables straightforward parameter recovery.}
\label{fig:loss}
\end{figure}

\paragraph{Coupling to differentiable hydro.}
Because \texttt{Ray-trax} is built from smooth interpolation/deposition primitives and JAX control flow, its Jacobians with respect to $(j,\kappa)$ are well defined.
When $(j,\kappa)$ comes from a differentiable hydrodynamics model (e.g., density, temperature, or ionization fractions produced by a differentiable PDE integrator; see \citep{2024arXiv241023093S}), \emph{the composed system remains differentiable}.
This opens the door to physics-constrained learning, gradient-based calibration of subgrid models, and end-to-end system identification where radiative observables inform hydrodynamic parameters.

\section{Discussion and conclusions}
We presented \textbf{\texttt{Ray-trax}}, a fast, \emph{time-dependent}, and fully differentiable ray tracer for on-the-fly emission--absorption RT on GPUs.
It operates directly on turbulent hydro snapshots, supports arbitrarily many frequency bins without code changes, and, despite a memory cost that scales as $\mathcal{O}(N_{\text{src}}\,N_{\text{cells}})$ when per-source fields are retained, remains highly efficient in practice due to full vectorization and sharding.
Analytical tests verify accuracy and convergence; turbulent applications demonstrate realism; and differentiability enables both inverse problems and seamless coupling to differentiable hydro while preserving gradients.
We view \texttt{Ray-trax} as a minimal, composable building block for modern astrophysical pipelines where RT can no longer be the bottleneck. The code is publicly available at \url{https://github.com/lorenzobranca/Ray-trax.git}.

\begin{section}{Acknowledgments}
This work is funded by the Carl-Zeiss-Stiftung through the NEXUS program. This work was supported by the Deutsche Forschungsgemeinschaft (DFG, German Research Foundation) under Germany’s Excellence Strategy EXC 2181/1 - 390900948 (the Heidelberg STRUCTURES Excellence Cluster). We acknowledge the usage of the AI-clusters Tom and Jerry funded by the Field of Focus 2 of Heidelberg University.
\end{section}

\bibliographystyle{plain}
\bibliography{main}

\appendix
\section{Notation}
\label{app:notation}

We summarize here the main symbols and terms used throughout the paper.

\paragraph{Geometry and coordinates.}
\begin{description}
\item[$\mathbf{x} = (x,y,z)$] Spatial position on a regular 3D Cartesian grid.
\item[$\hat{\mathbf{n}}$] Unit vector specifying a propagation direction on the unit sphere $\mathbb{S}^2$.
\item[$r_i = |\mathbf{x} - \mathbf{x}_i|$] Distance between position $\mathbf{x}$ and source position $\mathbf{x}_i$.
\end{description}

\paragraph{Radiation field.}
\begin{description}
\item[$I(\mathbf{x},\hat{\mathbf{n}},t)$] Specific intensity (monochromatic), i.e., radiant energy per unit time, area, solid angle, and frequency, at position $\mathbf{x}$, in direction $\hat{\mathbf{n}}$, and time $t$.
\item[$J(\mathbf{x},t)$] Angle-integrated (or accumulated) intensity at position $\mathbf{x}$ and time $t$, e.g.,
$J = \int I,\mathrm{d}\Omega$ or its discrete approximation via the sum over sampled directions.
\item[$J_{\mathrm{ref}}(\mathbf{x})$] Reference (analytical) solution for $J$ used in benchmarks.
\item[$I_{\mathrm{ref}}$] Reference intensity field used in inversion experiments.
\end{description}

\paragraph{Material properties and sources.}
\begin{description}
\item[$\kappa(\mathbf{x})$] Absorption coefficient (opacity) at position $\mathbf{x}$.
\item[$j(\mathbf{x})$] Emissivity at position $\mathbf{x}$ (source term in the emission--absorption equation).
\item[$\rho(\mathbf{x})$] Gas density field; in turbulent tests we set $\kappa(\mathbf{x}) \propto \rho(\mathbf{x})$.
\item[$L_i$] Luminosity (normalization) of the $i$-th point-like source in the analytical benchmark.
\item[$N_{\text{src}}$] Number of radiation sources.
\end{description}

\paragraph{Transport equation and marching scheme.}
\begin{description}
\item[$c$] Signal speed in the transport equation (speed of light or chosen radiation propagation speed).
\item[$t$] Physical time.
\item[$\Delta t$] Time-step size for the macroscopic evolution or light-travel horizon.
\item[$\Delta s$] Spatial step along a ray; we typically choose $\Delta s = c,\Delta t / N_s$ or an equivalent prescription.
\item[$N_s$] Number of substeps along each ray segment, $N_s = \lceil c,\Delta t / \Delta s \rceil$ (or user-specified).
\item[$k$] Substep index in the marching scheme.
\item[$\mathbf{x}k$] Ray position at substep $k$.
\item[$I_k$] Specific intensity at substep $k$ along a ray.
\item[$\tau_k$] Optical depth accumulated up to substep $k$,
$\tau{k+1} = \tau_k + \kappa(\mathbf{x}_k),\Delta s$.
\end{description}

\paragraph{Angular discretization.}
\begin{description}
\item[$N_\Omega$] Number of discrete propagation directions sampled on $\mathbb{S}^2$ (Fibonacci / golden-angle lattice).
\item[$4\pi/N_\Omega$] Solid angle weight used to approximate the integral over directions by a finite sum.
\end{description}

\paragraph{Discretization and complexity.}
\begin{description}
\item[$N_{\text{cells}}$] Total number of grid cells in the 3D domain.
\item[$\mathcal{O}(\cdot)$] Big-O notation for asymptotic computational or memory complexity.
\item[$\mathcal{O}(N_{\text{src}},N_\Omega,N_s)$] Work per time step: sources $\times$ directions $\times$ substeps.
\item[$\mathcal{O}(N_{\text{src}},N_{\text{cells}})$] Memory scaling when storing per-source fields before summation.
\end{description}

\paragraph{Inverse problems and differentiability.}
\begin{description}
\item[$A$] Scalar amplitude parameter (e.g., global scaling of a fixed emissivity template) recovered in gradient-based inversion.
\item[$e_0(\mathbf{x})$] Fixed emissivity template used in inversion experiments; the effective emissivity is $A,e_0(\mathbf{x})$.
\item[$\mathcal{L}$] Loss function used for calibration or inversion (e.g., mean squared error or relative error between $I$/$J$ and a reference).
\item[$\partial \mathcal{L}/\partial A$] Gradient of the loss with respect to $A$, obtained via automatic differentiation.
\end{description}

\paragraph{Implementation-level symbols (informal).}
\begin{description}
  \item[\texttt{j\textunderscore map}] Discrete grid representation of $j(\mathbf{x})$.
  \item[\texttt{kappa\textunderscore map}] Discrete grid representation of $\kappa(\mathbf{x})$.
  \item[\texttt{source\textunderscore pos}] Source position(s) in grid coordinates.
  \item[\texttt{num\textunderscore rays}] Implementation parameter corresponding to $N_\Omega$.
  \item[\texttt{step\textunderscore size}] Implementation parameter corresponding to $\Delta s$.
  \item[\texttt{radiation\textunderscore velocity}] Code parameter corresponding to $c$.
  \item[\texttt{time\textunderscore step}] Code parameter corresponding to $\Delta t$.
\end{description}

This notation table is intended to make the manuscript accessible to readers who are not specialists in radiative transfer while remaining consistent with standard usage in the field.
\end{document}